\documentclass[%
 reprint,
showpacs,
 amsmath,amssymb,
prb,
]{revtex4-1}
\usepackage{graphicx}% Include figure files
\usepackage{dcolumn}% Align table columns on decimal point
\usepackage{bm}% bold math
\usepackage{braket}
\usepackage{color}
\begin{document}
%\preprint{APS/123-QED}
%%%%%%%%%%%%%%%
%\title{Topological Classification and the realization of the flat-band zero-energy states at the dirty surface of a nodal superconductor}
\title{Symmetry conditions of a nodal superconductor for generating robust flat-band Andreev bound states at its dirty surface}
\author{Satoshi Ikegaya$^{1}$}
\author{Shingo Kobayashi$^{2,3}$}
\author{Yasuhiro Asano$^{1,4,5}$}%
\affiliation{
$^{1}$Department of Applied Physics,
Hokkaido University, Sapporo 060-8628, Japan\\
$^{2}$Department of Applied Physics,
Nagoya University, Nagoya 464-8603, Japan\\
$^{3}$Institute for Advanced Research, Nagoya University, Nagoya 464-8601, Japan\\
$^{4}$Center of Topological Science and Technology,
Hokkaido University, Sapporo 060-8628, Japan\\
$^{5}$Moscow Institute of Physics and Technology, 141700 Dolgoprudny, Russia}
%%%%%%%%%%%%%%%
\date{\today}
%%%%%%%%%%%%%%%
\begin{abstract}
We discuss the symmetry property of a nodal superconductor that hosts robust flat-band zero-energy states
at its surface under potential disorder. 
Such robust zero-energy states are known to induce the anomalous proximity effect 
in a dirty normal metal attached to a superconductor.
A recent study has shown that a topological 
index ${\cal N}_\mathrm{ZES}$ describes the number of zero-energy states at the dirty surface of 
a $p$-wave superconductor.
We generalize the theory to clarify the conditions 
required for a superconductor that enables ${\cal N}_\mathrm{ZES}\neq 0$. 
Our results show that ${\cal N}_\mathrm{ZES}\neq 0$
is realized in a topological material that belongs to either the BDI or CII class.
We also present two realistic Hamiltonians that result in ${\cal N}_\mathrm{ZES}\neq 0$.
\end{abstract}
%%%%%%%%%%%%%%%
\maketitle
%%%%%%%%%%%%%%%

%%%%%%%%%%%%%%%%%%%%%%%%%%%%%%%%%%%%%%%%%%%%
\section{Introduction}
%%%%%%%%%%%%%%%%%%%%%%%%%%%%%%%%%%%%%%%%%%%%
In the past decade, topologically nontrivial superconductors have attracted enormous attention
due to the existence of exotic bound states at their surfaces~\cite{kane_1,shou,sato_r}.
Early studies on this topic focused mainly on the topological phase
of a fully gapped superconductor listed in the tenfold topological classification~\cite{schnyder_0}.
The bulk-boundary correspondence suggests the equivalence between the number of surface bound states and
the absolute value of a topological invariant $\mathcal{Z}$ defined in the bulk states.
A unique physical consequence of such a topological superconductor might be the zero-bias conductance quantization 
at $G_{\mathrm{NS}}= (2e^2/h) |\mathcal{Z}|$ in a normal-metal/superconductor (NS) junction. 
In experiments, however, it is not easy to observe conductance quantization clearly
for the following reasons.
The invariant $\mathcal{Z}$ is usually limited to small numbers in real fully gapped superconductors,
whereas the number of propagating channels $N_c$ is much larger than unity 
in two- or three-dimensional NS junctions.
Therefore, the electric current passing through normal propagating channels
would smear the effects of resonant transmission through the topological bound states.

Today, superconductors characterized by
such unconventional pairing symmetry as spin-singlet $d$-wave and spin-triplet $p$-wave 
are considered to be topologically nontrivial although their gap functions have nodes on the Fermi surface
~\cite{yt95,ya04,sato_2,schnyder_5,buchholts,hu,yt95,ya04,yt10,yada,schnyder_1,schnyder_2,schnyder_3,alicea,you,si15,sato_12,law,deng,franz,kobayashi_2014,kobayashi_2015}.
The most striking feature of such a nodal superconductor is that it hosts flat-band zero-energy states (ZESs) at its clean surface. 
It has been well established that the conductance of an NS junction consisting of such an unconventional superconductor is quantized at 
$G_{\mathrm{NS}}= (2e^2/h) \mathcal{N}_{\mathrm{clean}}$~\cite{yt95,ya04,yt04,ya07,si15, si16}. 
Here $\mathcal{N}_{\mathrm{clean}}$ is the number of surface bound states at zero energy and is of the order of $N_c$.
In addition to zero-bias conductance quantization,
the fractional Josephson effect~\cite{yt96-1, barash, kwon, ya06,si16_2},
paramagnetic response at a surface~\cite{higashitani,walter,vorontsov,suzuki1,suzuki2,fogelstrom}
and the anomalies in heat transport~\cite{taddei_15}
are physical phenomena unique to a nodal superconductor.
Since $\mathcal{N}_{\mathrm{clean}}$ is the same order as $N_c$,
the effects of the surface bound states on the electromagnetic phenomena 
in a nodal superconductor can be more noticeable than those in a fully gapped topological superconductor.
However, these statements are true as long as the surface or the junction interface of the superconductor is sufficiently clean.
In experiments, potential disorder is inevitable at the surface of a superconductor
and may lift the degeneracy of the flat-band bound states at zero energy.
Actually, the flat-band ZESs at a surface of the $d$-wave superconductor are fragile under potential disorder~\cite{ya01,yt03}.
On the other hand, the flat-band ZESs of the $p$-wave superconductor are robust~\cite{yt04, ya06}.
The question is how to distinguish these two types of nodal superconductors.

A key theoretical method with which to solve the problem is called dimensional reduction,
and it is a useful theoretical tool for characterizing a nodal superconductor topologically.
In a $d$-dimensional nodal superconductor, 
we can still find a fully gapped one-dimensional partial Brillouin zone
by fixing the ($d-1$)-dimensional momentum at a certain point (say $k$).
In such a one-dimensional Brillouin zone, it is possible to define a winding number $w(k)$ in terms of the wave 
function of the occupied states below the gap~\cite{sato_2,schnyder_5}.
A nonzero winding number $w(k)$ suggests $|w(k)|$-fold degenerate ZESs at the surface for each $k$.
Therefore, $\mathcal{N}_{\mathrm{clean}}= \sum_{k}| w(k)|$ 
describes the number of ZESs at the clean surface of a nodal superconductor.
In contrast to the topological invariant of a fully gapped topological superconductor, 
$\mathcal{N}_{\mathrm{clean}}$ cannot predict the number of ZESs at a dirty surface~\cite{schnyder_6}.
With the dimensional reduction, translational symmetry is necessary to define the winding number 
in a one-dimensional Brillouin zone.
However, such partial Brillouin zones itself are not well defined at all 
because the momentum $k$ is no longer a good quantum number under potential disorder. 
Nevertheless, two of the present authors have shown that
an alternative index ${\cal N}_\mathrm{ZES}=\sum_k w(k)$ describes
the number of ZESs at a dirty surface~\cite{si17}.
In other words, ${\cal N}_\mathrm{ZES}$ represents the bulk-boundary correspondence of a 
nodal superconductor in the dirty case. 
Moreover, a nodal superconductor with ${\cal N}_\mathrm{ZES}\neq 0$
is known to induce the anomalous proximity effect in dirty proximity structures such as
conductance quantization at $G_{\mathrm{NS}}= (2e^2/h) |\mathcal{N}_\mathrm{ZES}|$ in a dirty NS junction~\cite{yt04}, 
the fractional Josephson effect in a dirty Josephson junction~\cite{ya06},
and the paramagnetic Meissner response at a dirty surface of a superconductor~\cite{suzuki2}.  
To our knowledge, ${\cal N}_\mathrm{ZES}$ becomes nonzero in several nodal superconductors
characterized by spin-triplet $p$- and $f$-wave pairing symmetries. Namely spin-triplet $p$- and $f$-wave superconductivity 
is a sufficient condition for ${\cal N}_\mathrm{ZES}\neq 0$. 
However, we have never known any necessary conditions for ${\cal N}_\mathrm{ZES}\neq 0$.
Such conditions would provide us with a design guide for topologically nontrivial artificial superconductors, 
which may be realized by applying the fabrication technique to existing materials.
This paper will clarify the necessary conditions for ${\cal N}_\mathrm{ZES}\neq 0$.

In this paper,
we first study the relationship between
the symmetry class of the Bogoliubov-de Gennes (BdG) Hamiltonian 
and the possibility of a nonzero index ${\cal N}_\mathrm{ZES}$.
Within the tenfold topological classification,
the classes BDI, CII, DIII, and CI are the target symmetry classes of this paper
because a nodal superconductor belonging to these symmetry classes
is able to host flat-band ZESs at their clean surfaces.
We find that ${\cal N}_\mathrm{ZES}=0$ identically in classes DIII and CI,
whereas ${\cal N}_\mathrm{ZES}\neq 0$ is realized in classes BDI and CII.
The results are summarized in Table.~\ref{tab:result}.
On the basis of this conclusion,
we also seek practical examples of the BdG Hamiltonian in class BDI.
As a result, we find two realistic models, which describe
a Dresselhaus [110] superconductor~\cite{you,si15} and 
a helical $p$-wave superconductor with an in-plane magnetic field~\cite{sato_12,law}.
Thus we conclude that these superconductors host
$|{\cal N}_\mathrm{ZES}|$-fold degenerate ZESs at their dirty surfaces.

This paper is organized as follows.
In Sec.~\ref{sec:2}, we discuss the possible symmetry class of a nodal superconductor
that possesses the nonzero index ${\cal N}_\mathrm{ZES}$.
On the basis of the general conclusion in Sec.~\ref{sec:2},
we present realistic models of a nodal superconductor in class BDI in Sec.~\ref{sec:3}.
We summarize this paper in Sec.~\ref{sec:4}.

%%%%%%%%%%%%%%%%%%%%%%%%%%%%%%%%%%%%%%%%%%%%%%%%%%
\section{Symmetry class and Topological index}
\label{sec:2}
%%%%%%%%%%%%%%%%%%%%%%%%%%%%%%%%%%%%%%%%%%%%%%%%%%

%%%%%%%%%%%%%%%%%%%%%%%%%%%%%%%%%%%%%%%%%%%%%%%%%%
\subsection{Preliminary}
%%%%%%%%%%%%%%%%%%%%%%%%%%%%%%%%%%%%%%%%%%%%%%%%%%
First, we briefly review the topological property of a time-reversal invariant nodal superconductor.
The Bogoliubov-de Gennes (BdG) Hamiltonian in momentum space is generally given by
\begin{align}
{\cal H}(\boldsymbol{k}) = \left[ \begin{array}{cc}
h (\boldsymbol{k}) &
\Delta (\boldsymbol{k}) \\
- \Delta^{\ast} (-\boldsymbol{k}) &
-h^{\ast} (-\boldsymbol{k}) \\
\end{array}\right],
\label{eq:bdg}
\end{align}
where $h (\boldsymbol{k})$ denotes the $N \times N$ normal state Hamiltonian of an electron,
$\Delta (\boldsymbol{k})$ is the $N \times N$ pair potential,
and $N$ represents the number of degrees of freedom such as spins and conduction bands.
Time-reversal symmetry (TRS) and particle-hole symmetry (PHS) of ${\cal H}(\boldsymbol{k})$ are represented by
\begin{align}
&{\cal T} \; {\cal H}(\boldsymbol{k}) \; {\cal T}^{-1} = {\cal H}(-\boldsymbol{k}), 
\quad {\cal T} = {\cal U}_{\cal T} {\cal K}, \quad {\cal T}^2 = \pm 1,
\label{eq:trs}\\
&{\cal C} \; {\cal H}(\boldsymbol{k}) \; {\cal C}^{-1} = - {\cal H}(-\boldsymbol{k}),
\quad {\cal C} = {\cal U}_{\cal C} {\cal K}, \quad {\cal C}^2 = \pm 1,
\label{eq:phs}
\end{align}
where ${\cal U}_{\cal T}$ and ${\cal U}_{\cal C}$ are $2N \times 2N$ unitary operators,
and ${\cal K}$ is the complex conjugation operator.
In terms of the signs of ${\cal T}^2$ and ${\cal C}^2$,
we can classify the present BdG Hamiltonian into four symmetry classes: BDI, CII, DIII, and CI~\cite{schnyder_0}. 
The values of $({\cal T}^2, {\cal C}^2)$ in these classes are summarized in Table~\ref{tab:result}.

When a BdG Hamiltonian belongs to one of these symmetry classes,
it is possible to define chiral symmetry (CS) of the Hamiltonian by
\begin{align}
{\cal S} \; {\cal H}(\boldsymbol{k}) \; {\cal S}^{-1} = - {\cal H}(\boldsymbol{k}),
\quad {\cal S} = e^{i \alpha} {\cal T}{\cal C},
\label{eq:cs}
\end{align}
where ${\cal S}$ is a unitary operator and $\alpha$ is an arbitrary real number.
The commutation relation $[{\cal S}^2, {\cal H}(\boldsymbol{k})]=0$
holds for any Hamiltonians preserving chiral symmetry.
As a result, ${\cal S}^2$ is proportional to the identity operator as ${\cal S}^2=e^{i \beta}$.
Phase $\beta$ can be removed by
choosing $\alpha$ in an appropriate way.
Thus, in the following, we assume ${\cal S}^2=+1$ without loss of generality.

In the superconductor under consideration, the pair potential has nodes on the Fermi surface.
Therefore, it is impossible to define a topological invariant by using the wave functions of
the entire Brillouin zone. 
In a three- (two-) dimensional case,
we assume that the pair potential has line (point) nodes on the Fermi surface. 
The nodal point 
$\boldsymbol{k}_0$ satisfies $\mathrm{det}[{\cal H}(\boldsymbol{k}_0)]=0$.
Even in the presence of the nodes,
it is still possible to define a one-dimensional partial Brillouin zone
by fixing the $(d-1)$-dimensional momentum $\boldsymbol{k}_\parallel$ at a certain point.
When the pair potential in such a partial Brillouin zone is fully gapped,
we can define the one-dimensional winding number as
\begin{align}
w(\boldsymbol{k}_\parallel) = \frac{i}{4 \pi}
\int dk_\perp \mathrm{Tr}
[ {\cal S} {\cal H}^{-1}(\boldsymbol{k})
\partial_{k_\perp} {\cal H}(\boldsymbol{k}) ],
\label{eq:wind}
\end{align}
where $k_\perp$ is the momentum in a one-dimensional Brillouin zone.
The winding number $w(\boldsymbol{k}_\parallel)$ cannot be defined when the integral path along
$k_\perp$ in Eq.~(\ref{eq:wind}) intersects the gap nodes $\boldsymbol{k}_0$.

When $w(\boldsymbol{k}_\parallel)$ is nonzero at $\boldsymbol{k}_\parallel$,
according to the bulk-boundary correspondence, $|w(\boldsymbol{k}_\parallel)|$-fold degenerate 
ZESs are expected at a clean surface parallel to $\boldsymbol{k}_\parallel$.
Thus, the total number of ZESs at a clean surface is given by
\begin{align}
{\cal N}_\mathrm{clean} =
{\sum_{\boldsymbol{k}_{\parallel}}}^{\prime} |w(\boldsymbol{k}_{\parallel})|,
\end{align}
where ${\sum_{\boldsymbol{k}_{\parallel}}}^{\prime}$ denotes
a summation over $\boldsymbol{k}_{\parallel}$ excluding the nodal points.
Such highly degenerate surface bound states are called flat-band ZESs because
the energy dispersion is independent of $\boldsymbol{k}_{\parallel}$.

Next, we focus on flat-band ZESs at the dirty surface of a nodal superconductor.
The surface is located at $x_\perp = 0$ and the nodal superconductor occupies $x_\perp \geq 0$.
The potential disorder in the bulk region strongly suppresses unconventional superconductivity.
Thus, we assume that the potential disorders exist only near the surface $x_\perp \ll \xi_\mathrm{S}$,
where $\xi_\mathrm{S}$ represents the superconducting coherence length.
The non-magnetic random potential $V(\boldsymbol{r})$ preserves TRS and PHS as
\begin{align}
&{\cal T} \; V(\boldsymbol{r}) \; {\cal T}^{-1} = V(\boldsymbol{r}),\\
&{\cal C} \; V(\boldsymbol{r}) \; {\cal C}^{-1} = - V(\boldsymbol{r}),
\end{align}
where $V(\boldsymbol{r})$ is finite only for $x_\perp \ll \xi_\mathrm{S}$.
In the presence of potential disorders, 
the winding number $w(\boldsymbol{k}_\parallel)$ is not well defined
because the momentum $\boldsymbol{k}_{\parallel}$ is no longer a good quantum number 
in the absence of translational symmetry.
This implies that $w(\boldsymbol{k}_\parallel)$ cannot straightforwardly 
predict the number of the ZESs at a dirty surface.
Nevertheless, it is possible to characterize the flat-band ZESs at a dirty surface
by using an alternative index~\cite{si17} 
\begin{align}
{\cal N}_\mathrm{ZES} 
= {\sum_{\boldsymbol{k}_{\parallel}}}^{\prime} w(\boldsymbol{k}_{\parallel}).
\label{eq:n_zes}
\end{align}
The absolute value of the index ${\cal N}_\mathrm{ZES}$
coincides with the number of ZESs at the dirty surface
as long as chiral symmetry is preserved (See also section II in Ref.~[\onlinecite{si17}]).
In what follows,
we study the relationship between the symmetry class of the Hamiltonian
and the realization of the nonzero index ${\cal N}_\mathrm{ZES}$.

%%%%%%%%%%%%%%%%%%%%%%%%%%%%%%%%%%%%%%%%%%%%%%%%%%
\subsection{Realization of nonzero topological index}
%%%%%%%%%%%%%%%%%%%%%%%%%%%%%%%%%%%%%%%%%%%%%%%%%%
As shown in Appendix~\ref{sec:com}, 
the commutation relations for the symmetry operators depend on
the symmetry class of the Hamiltonian as follows
\begin{align}
&[{\cal S}, {\cal T}] = [{\cal S}, {\cal C}] = 0 \qquad \text{for \; BDI and CII}, \label{eq:cs1} \\
&\left\{{\cal S}, {\cal T}\right\} = \left\{{\cal S}, {\cal T}\right\} = 0 \qquad \text{for \; DIII and CI}. \label{eq:cs2}
\end{align}
From these commutation relations, we obtain
\begin{align}
&{\cal T}^{-1}\; {\cal S} \;{\cal T} = {\cal U}_{\cal T}^{\dagger} \; {\cal S}^{\ast} \; {\cal U}_{\cal T} = \eta \,{\cal S}, \label{eq:usu}\\
&{\cal C}^{-1} \; {\cal S} \; {\cal C} = {\cal U}_{\cal C}^{\dagger} \; {\cal S}^{\ast} \; {\cal U}_{\cal C} = \eta \, {\cal S}, \label{eq:usu2}\\
&\eta =\left\{ \begin{array}{ll} +1\; \text{for \; BDI and CII} \\ -1\; \text{for \; DIII and CI.} \end{array}\right.
\end{align}
By taking account of Eqs.~(\ref{eq:usu}) and (\ref{eq:usu2}),
 the complex conjugation of the winding number~\cite{yamakage} is calculated as follows
\begin{widetext}
\begin{align}
\left\{ w(\boldsymbol{k}_{\parallel})\right\}^{\ast}
&= - \frac{i}{4 \pi} \int d k_{\perp} \mathrm{Tr} \left[
{\cal S}^{\ast} \left\{ {\cal H}^{-1} (\boldsymbol{k}) \right\}^{\ast}
\partial_{k_{\perp}} {\cal H}^{\ast} (\boldsymbol{k}) \right] \nonumber\\
&= - \frac{i}{4 \pi} \int d k_{\perp} \mathrm{Tr} \left[
{\cal S}^{\ast}
\left\{ {\cal U}_{\mathrm{\Lambda}} \; {\cal H}^{-1} (- \boldsymbol{k}) \; {\cal U}_{\Lambda}^{\dagger} \right\}
\; \partial_{k_{\perp}} 
\left\{ {\cal U}_{\Lambda} \; {\cal H} (- \boldsymbol{k}) \; {\cal U}_{\Lambda}^{\dagger} \right\} \right] \nonumber\\
&=- \frac{i}{4 \pi} \int d k_{\perp} \mathrm{Tr} \left[
\left\{ {\cal U}_{\Lambda}^{\dagger}\;  {\cal S}^{\ast} \; {\cal U}_{\Lambda} \right\}
{\cal H}^{-1} (- \boldsymbol{k})
\; \partial_{k_{\perp}} {\cal H} (- \boldsymbol{k})  \right] \nonumber\\
&=\frac{i}{4 \pi} \int d k_{\perp} \mathrm{Tr} \left[
\left( \eta {\cal S} \right) {\cal H}^{-1} (k_\perp, - \boldsymbol{k}_\parallel)
\; \partial_{k_{\perp}} {\cal H} (k_\perp, - \boldsymbol{k}_\parallel) \right] \nonumber\\
&= \eta \, w(- \boldsymbol{k}_{\parallel}), \label{eq:wind_c}
\end{align}
\end{widetext}
where $\Lambda={\cal T}$ or ${\cal C}$.
In the second line of Eq.~(\ref{eq:wind_c}), we use the relations
\begin{align}
&{\cal H}^{\ast} (\boldsymbol{k}) = {\cal U}_{\cal T} \;  {\cal H}(- \boldsymbol{k}) \; {\cal U}_{\cal T}^{\dagger}, \label{eq:trs2}\\
&{\cal H}^{\ast} (\boldsymbol{k}) = - {\cal U}_{\cal C} \; {\cal H}(- \boldsymbol{k}) \; {\cal U}_{\cal C}^{\dagger}, \label{eq:phs2}
\end{align}
which are equivalent to TRS in Eq.~(\ref{eq:trs}) and PHS in Eq.~(\ref{eq:phs}), respectively.
Since $w(\boldsymbol{k}_{\parallel})$ is a real integer number,
we finally obtain an important relation
\begin{align}
w(\boldsymbol{k}_{\parallel}) =  \eta \, w(- \boldsymbol{k}_{\parallel}).
\label{eq:w-w}
\end{align}
From Eq.~(\ref{eq:w-w}),
we find that the winding number for classes DIII and CI (i.e., $\eta = -1$)
is an odd function of $\boldsymbol{k}_{\parallel}$.
Therefore, the index ${\cal N}_\mathrm{ZES}$ in Eq.~(\ref{eq:n_zes}) becomes identically zero.
This implies the absence of zero-energy states at the dirty surfaces of DIII and CI nodal superconductors.
On the other hand, the winding number for classes BDI and CII (i.e., $\eta = +1$)
is an even function of $\boldsymbol{k}_{\parallel}$.
Therefore, ${\cal N}_\mathrm{ZES} \neq 0$ is possible in these symmetry classes,
which means that degenerate zero-energy states exist at the dirty surface.
We summarize the results in Table~\ref{tab:result}.
At a clean surface,
flat-band ZESs are expected irrespective of the symmetry classes of a nodal superconductor
(See the fifth column of Table~\ref{tab:result}).
However, at a realistic dirty surface,
the presence or absence of the flat-band ZESs depends on the symmetry class of the superconductor
(See the sixth column of Table~\ref{tab:result}).
Namely, only the BDI or CII nodal superconductor has the potential
to host degenerate ZESs at its dirty surface.
This is the main conclusion of this paper.

The BdG Hamiltonian in class CI describes a spin-singlet superconductor~\cite{schnyder_0}.
Therefore, the flat-band ZESs of a spin-singlet $d_{xy}$-wave superconductor
are fragile against potential disorder~\cite{ya01,yt03}.
Although several noncentrosymmetric superconductors
have flat-band ZESs at their clean surface~\cite{yt10,yada,schnyder_1,schnyder_2,schnyder_3},
the potential disorder completely lift the degeneracy at zero energy~\cite{schnyder_6, si17}
because the noncentrosymmetric superconductors belong to class DIII.
The BdG Hamiltonian of a spin-triplet superconductor preserving spin-rotation symmetry
belongs to class BDI~\cite{si17}.
Actually, the flat-band ZESs of the spin-triplet $p_x$-wave superconductor can retain their high degree of degeneracy
even in the presence of the potential disorder~\cite{yt04, ya06, si17}.
In the following section, we investigate other examples of nodal superconductors
that host robust flat-band ZESs under potential disorder.
Unfortunately, we cannot find a specific model of a nodal superconductor in class CII.
Even so, we demonstrate two practical models of topologically nontrivial nodal superconductors
belonging to class BDI.

%-------------------------------------------------------------------------------------------------------------------
\begin{table}
\caption{\label{tab:result}
Relationship between the symmetry class of a nodal superconductor
and the number of flat-band zero-energy states (ZESs) at its surface.
The first column lists the relevant symmetry classes of the nodal superconductor.
The second and third columns indicate the sign of ${\cal T}^2$ and ${\cal C}^2$, respectively.
The fourth column indicates the presence of chiral symmetry by ${\cal S}^2=+1$.
The fifth column represents the number of flat-band ZESs
at a clean surface of a nodal superconductor ${\cal N}_\mathrm{clean} =
{\sum_{\boldsymbol{k}_{\parallel}}}^{\prime} |w(\boldsymbol{k}_{\parallel})|$.
The sixth column denotes
the number of the ZESs at a dirty surface of a nodal superconductor
evaluated by the index
${\cal N}_\mathrm{ZES} =
{\sum_{\boldsymbol{k}_{\parallel}}}^{\prime} w(\boldsymbol{k}_{\parallel}).$}
\begin{ruledtabular}
\begin{tabular}{cccccc}
& \textrm{TRS} & \textrm{PHS} & \textrm{CS}
&\textrm{clean} & \textrm{dirty}\\ \hline
\textrm{BDI} & $+1$ & $+1$ & $+1$ & ${\cal N}_\mathrm{clean}$ & $|{\cal N}_\mathrm{ZES}|$ \\
\textrm{CII} & $-1$ & $-1$ & $+1$ & ${\cal N}_\mathrm{clean}$& $|{\cal N}_\mathrm{ZES}|$\\
\textrm{DIII} & $-1$ & $+1$ & $+1$ & ${\cal N}_\mathrm{clean}$  & $0$  \\
\textrm{CI} & $+1$ & $-1$ & $+1$ & ${\cal N}_\mathrm{clean}$& $0$ \\
\end{tabular}
\end{ruledtabular}
\end{table}
%-------------------------------------------------------------------------------------------------------------------

%%%%%%%%%%%%%%%%%%%%%%%%%%%%%%%%%%%%%%%%%%%%
\section{Nodal superconductors with the nonzero topological index}
\label{sec:3}
%%%%%%%%%%%%%%%%%%%%%%%%%%%%%%%%%%%%%%%%%%%%

%%%%%%%%%%%%%%%%%%%%%%%%%%%%%%%%%%%%%%%%%%%%%%
\subsection{BdG Hamiltonian in the single-band model}
%%%%%%%%%%%%%%%%%%%%%%%%%%%%%%%%%%%%%%%%%%%%%%

In this paper, we restrict our discussion to single-band superconductors belonging to class BDI.
The BdG Hamiltonian in the single-band model is generally given by
\begin{align}
&\check{H}(\boldsymbol{k}) = \left[ \begin{array}{cc}
\hat{h}(\boldsymbol{k}) &
\hat{\Delta} (\boldsymbol{k}) \\
-\hat{\Delta}^{\ast} (-\boldsymbol{k}) &
-\hat{h}(\boldsymbol{k})^{\ast} (-\boldsymbol{k}) \\
\end{array}\right], \label{eq:bdg_single}\\
&\hat{h}(\boldsymbol{k})=
\varepsilon(\boldsymbol{k}) \sigma_0
+ \boldsymbol{g}(\boldsymbol{k}) \cdot \boldsymbol{\sigma}
+ \boldsymbol{V} \cdot \boldsymbol{\sigma},\label{eq:g_hn} \\
&\hat{\Delta} (\boldsymbol{k})
=\left[
\psi(\boldsymbol{k}) + \boldsymbol{d}(\boldsymbol{k}) \cdot \boldsymbol{\sigma}
\right] ( i \sigma_2),
\label{eq:g_pair}\\
&\varepsilon(\boldsymbol{k})=
\frac{\hbar^2 \boldsymbol{k}^2}{2m} - \mu_\mathrm{F},
\end{align}
where $\sigma_0$ is the $2 \times 2$ unit matrix,
$m$ denotes the effective mass of an electron,
and $\mu_\mathrm{F}$ is the chemical potential.
The spin-orbit coupling potential is given by $\boldsymbol{g}(\boldsymbol{k})= -\boldsymbol{g}(-\boldsymbol{k})$.
The Zeeman potential induced by an external magnetic field is denoted by $\boldsymbol{V}$.
The pair potential of a spin-singlet even-parity pairing order and that of a spin-triplet odd-parity 
pairing order are represented by $\psi(\boldsymbol{k})=\psi(-\boldsymbol{k})$
and $\boldsymbol{d}(\boldsymbol{k})=-\boldsymbol{d}(-\boldsymbol{k})$, respectively.
In what follows, we assume the time-reversal invariant pairing orders which satisfy
$\psi^{\ast}(\boldsymbol{k}) = \psi(\boldsymbol{k})$
and $\boldsymbol{d}^{\ast}(\boldsymbol{k})=\boldsymbol{d}(\boldsymbol{k})$.
The BdG Hamiltonian preserves PHS intrinsically as 
\begin{align}
\check{C}_+ \; \check{H}(\boldsymbol{k}) \; \check{C}_+^{-1}=-\check{H}(-\boldsymbol{k}),\quad
\check{C}_+  = \left[ \begin{array}{cc}
0 & \sigma_0 \\ \sigma_0 & 0 \\ \end{array}\right] {\cal K},
\end{align}
where $\check{C}_+^2=+1$.
For spinful fermionic systems, the TRS operator is generally defined by
\begin{align}
\check{T}_-= \left[ \begin{array}{cc}
i \sigma_2  & 0 \\ 0 &  i \sigma_2  \\ \end{array}\right] {\cal K}
\end{align}
obeying $\check{T}_-^2=-1$.
In the absence of the Zeeman potential (i.e., $\boldsymbol{V}=0$),
the BdG Hamiltonian $\check{H}(\boldsymbol{k})$ satisfies
$\check{T}_- \; \check{H}(\boldsymbol{k}) \; \check{T}_-^{-1}=\check{H}(-\boldsymbol{k})$,
which represents TRS of the BdG Hamiltonian.
On the basis of the results in Sec.~\ref{sec:2}, however,
the index ${\cal N}_\mathrm{ZES}$ defined by using
the chiral symmetry operator $\check{S}^{\prime}=-i \check{T}_- \check{C}_+$ becomes identically zero.
Alternatively,
we assume that the BdG Hamiltonian $\check{H}(\boldsymbol{k})$ satisfies
\begin{align}
\check{T}_+ \; \check{H}(\boldsymbol{k}) \; \check{T}_+^{-1}=\check{H}(-\boldsymbol{k}),
\end{align}
where $\check{T}_+$ is a $4 \times 4$ anti-unitary operator satisfying $\check{T}_+^2=+1$.
In the single-band model, 
the anti-unitary operator $\check{T}_+$ is defined
by combining the original TRS operator $\check{T}_-$ and an unitary operator $\check{R}$ as
\begin{align}
&\check{T}_+ = \check{R} \; \check{T}_-
= \left[ \begin{array}{cc}
\hat{r}( i \sigma_2 ) & 0 \\ 0 & \hat{r}^{\ast} ( i \sigma_2 ) \\ \end{array}\right] {\cal K}, 
\label{eq:trs_+1}\\
&\check{T}_-= \left[ \begin{array}{cc}
i \sigma_2  & 0 \\ 0 &  i \sigma_2  \\ \end{array}\right] {\cal K}, \quad
\check{R}= \left[ \begin{array}{cc}
\hat{r}  & 0 \\ 0 &  \hat{r}^{\ast}  \\ \end{array}\right],
\end{align}
where $\hat{r}$ is a $2 \times 2$ unitary operator.
To satisfy the condition $\check{T}_+^2=+1$,
the form of the unitary operator $\hat{r}$ is restricted as
\begin{align}
&\hat{r} = - i e^{i \gamma/2} \boldsymbol{n} \cdot \boldsymbol{\sigma}, 
\end{align}
where $\gamma$ is an arbitrary real number,
and $\boldsymbol{n}$ is a unit vector in an arbitrary direction in spin space (See also Appendix~\ref{sec:aut}).
As shown in Appendix~\ref{sec:uni}, it is possible to choose $\boldsymbol{n}$ in the specific direction 
because all the BdG Hamiltonians satisfy Eq.~(\ref{eq:trs_+1})
are always unitary equivalent to one another. In this paper, therefore, we choose $\boldsymbol{n}$ being in the third 
spin direction and consider 
\begin{align}
&\check{T}_z \; \check{H}(\boldsymbol{k}) \; \check{T}_z^{-1} =\check{H}(-\boldsymbol{k}),
\label{eq:trs_nb}\\
&\check{T}_z =  \check{R}_z \; \check{T}_-
= \left[ \begin{array}{cc}
-i e^{i \gamma/2} \sigma_1  & 0 \\ 0 & i e^{-i \gamma/2} \sigma_1  \\ \end{array}\right] {\cal K},
\label{eq:trs_+2}\\
&\check{R}_z= \left[ \begin{array}{cc}
\hat{r}_z  & 0 \\ 0 &  \hat{r}_z^{\ast}  \\ \end{array}\right], \quad
\hat{r}_z= - i e^{i \gamma/2} \sigma_3,
\end{align}
in what follows.

In Eq.~(\ref{eq:trs_nb}),
the normal state Hamiltonian $\hat{h}(\boldsymbol{k})$
and the pair potential $\hat{\Delta}(\boldsymbol{k})$ respectively obey the relations
\begin{align}
&\hat{T}_z \; \hat{h}(\boldsymbol{k}) \; \hat{T}_z^{\dagger} = \hat{h}(-\boldsymbol{k}),
\label{eq:bdg_n}\\
&\hat{T}_z \; \hat{\Delta}(\boldsymbol{k}) \; \hat{T}_z^{\mathrm{T}} = \hat{\Delta}(-\boldsymbol{k}),
\label{eq:bdg_p}
\end{align}
where $\hat{T}_z = -i e^{i \gamma/2} \sigma_1 {\cal K}$,
and $\mathrm{T}$ means a transpose of a matrix.
The normal state Hamiltonian in Eq.~(\ref{eq:g_hn}) is transformed into
\begin{align}
\hat{T}_z \; \hat{h}(\boldsymbol{k}) \; \hat{T}_z^{\dagger} =& \varepsilon(\boldsymbol{k}) \sigma_0
- g_3(\boldsymbol{k}) \sigma_3 + \sum_{j=1,2} V_j \sigma_j \nonumber\\
&+ \sum_{j=1,2} g_j (\boldsymbol{k}) \sigma_j - V_3 \sigma_3.
\end{align}
To satisfy the equation~(\ref{eq:bdg_n}), 
the normal Hamiltonian should have a form
\begin{align}
&\hat{h}_\mathrm{BDI}(\boldsymbol{k})= \varepsilon(\boldsymbol{k}) \sigma_0
+ g_3(\boldsymbol{k}) \sigma_3
+ \sum_{j=1,2} V_j \sigma_j.
\end{align}
The pair potential in Eq.~(\ref{eq:g_pair}) is transformed into
\begin{align}
\hat{T}_z \; \hat{\Delta}(\boldsymbol{k}) \; \hat{T}_z^{\mathrm{T}} 
= & e^{i \gamma} \left[
\psi(\boldsymbol{k}) - d_3(\boldsymbol{k}) \sigma_3 \right] (i \sigma_2) \nonumber\\
&+ e^{i \gamma} \sum_{j=1,2} d_j (\boldsymbol{k}) \sigma_j (i \sigma_2).
\end{align}
There are two possible choices of $\hat{\Delta} (\boldsymbol{k})$ and $\gamma$
 to satisfy the equation~(\ref{eq:bdg_p}).
The first choice is
\begin{align}
\hat{\Delta}_1 (\boldsymbol{k})
=\left[ \psi(\boldsymbol{k}) + d_3 (\boldsymbol{k}) \sigma_3\right] (i \sigma_2),
\end{align}
with setting $e^{i \gamma}=+1$.
The second one is
\begin{align}
\hat{\Delta}_2 (\boldsymbol{k})
= \sum_{j=1,2} d_j (\boldsymbol{k})\sigma_j  (i \sigma_2),
\end{align}
with setting $e^{i \gamma}=-1$.
As a consequence,
the BdG Hamiltonian belonging to class BDI can be represented as
\begin{align}
&\check{H}_\mathrm{BDI}(\boldsymbol{k}) = \left[ \begin{array}{cc}
\hat{h}_\mathrm{BDI}(\boldsymbol{k}) &
\hat{\Delta} _\lambda(\boldsymbol{k}) \\
-\hat{\Delta}_\lambda^{\ast} (-\boldsymbol{k}) &
-\hat{h}_\mathrm{BDI}^{\ast} (-\boldsymbol{k}) \\
\end{array}\right], \label{eq:bdg_bdi}\\
&\hat{h}_\mathrm{BDI}(\boldsymbol{k})= \varepsilon(\boldsymbol{k}) \sigma_0
+ g_3(\boldsymbol{k}) \sigma_3 + \sum_{j=1,2} V_j \sigma_j,\\
&\hat{\Delta}_1 (\boldsymbol{k})
=\left[ \psi(\boldsymbol{k}) + d_3 (\boldsymbol{k}) \sigma_3\right] i \sigma_2,\\
&\hat{\Delta}_2 (\boldsymbol{k}) = \sum_{j=1,2} d_j (\boldsymbol{k})\sigma_j  i \sigma_2.
\end{align}
When we chose the pair potential $\hat{\Delta}_1 (\boldsymbol{k})$,
the corresponding TRS operator is given by
\begin{align}
&\check{T}_{+,1} = \check{\theta}_z \; \check{T}_-
= \left[ \begin{array}{cc}
- i \sigma_1 & 0 \\ 0 & i \sigma_1 \\ \end{array}\right] {\cal K},
\label{eq:t+1}\\
&\check{\theta}_z = \left[ \begin{array}{cc}
- i \sigma_3 & 0 \\ 0 & i \sigma_3 \\ \end{array}\right],
\end{align}
where the unitary operator $\check{\theta}_z$ physically means the spin-rotation around the $z$-axis.
When we chose the pair potential $\hat{\Delta}_2(\boldsymbol{k})$,
on the other hand,
the corresponding TRS operator becomes
\begin{align}
&\check{T}_{+,2} = \check{\chi} (\pi) \; \check{\theta}_z \; \check{T}_-
= \left[ \begin{array}{cc}
\sigma_1 & 0 \\ 0 & \sigma_1 \\ \end{array}\right] {\cal K},
\label{eq:t+2}\\
&\check{\chi} (\pi) = \left[ \begin{array}{cc}
e^{i \pi/2}\sigma_0 & 0 \\ 0 & e^{- i \pi/2}\sigma_0 \\ \end{array}\right],
\end{align}
where $\check{\chi} (\pi)$ represents the gauge transformation by $\pi$.
In the following subsections,
we discuss realistic two examples of nodal superconductors whose 
BdG Hamiltonians satisfy Eq.~(\ref{eq:bdg_bdi}).

%%%%%%%%%%%%%%%%%%%%%%%%%%%%%%%%%%%%%%%%%%%%%%
\subsection{Dresselhaus [110] superconductor}
%%%%%%%%%%%%%%%%%%%%%%%%%%%%%%%%%%%%%%%%%%%%%%

%------------------------------------------------------------------------
\begin{figure}[htbp]
\begin{center}
\includegraphics[width=0.3\textwidth]{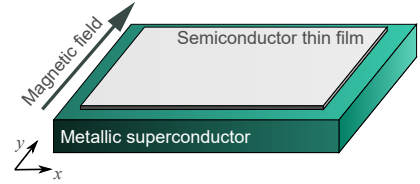}
\caption{(Color online)
Schematic image of a Dresselhaus[110] superconductor.
}
\label{fig:model_d}
\end{center}
\end{figure}
%------------------------------------------------------------------------

The first example may be an artificial superconducting hybrid, where 
a semiconductor thin film with the strong Dresselhaus [110] spin-orbit coupling is
fabricated on a metallic superconductor~\cite{you,si15} as shown in Fig.~\ref{fig:model_d}.
The semiconductor thin film is superconductive due to the proximity-effect induced 
$s$-wave pair potential.
We also apply an in-plane magnetic field which induces
the Zeeman potential on the thin film.
Such superconducting film is described by the BdG Hamiltonian
\begin{align}
&\check{H}_\mathrm{D}(\boldsymbol{k}) = \left[ \begin{array}{cc}
\hat{h}_\mathrm{D}(\boldsymbol{k}) &
\hat{\Delta}_\mathrm{D}(\boldsymbol{k}) \\
- \hat{\Delta}_\mathrm{D}^{\ast}(-\boldsymbol{k}) &
-\hat{h}_\mathrm{D}^{\ast} (-\boldsymbol{k}) \\
\end{array}\right], \label{eq:bdg_single}\\
&\hat{h}_\mathrm{D}(\boldsymbol{k})=
\varepsilon(\boldsymbol{k}) \sigma_0 + \beta k_x \sigma_3 +  \sum_{j=1,2} V_j \sigma_j,\\
&\hat{\Delta}_\mathrm{D}(\boldsymbol{k}) = i \Delta_s \sigma_2,
\end{align}
where $\beta$ is the strength of the Dresselhaus[110] spin-orbit coupling 
and $\Delta_s$ represents the amplitude of the proximity induced $s$-wave pair potential.
The BdG Hamiltonian in Eq.~(\ref{eq:bdg_single}) satisfies
\begin{align}
&\check{T}_{+,1} \; \check{H}_\mathrm{D}(\boldsymbol{k}) \; \check{T}_{+,1}^{-1}
=\check{H}_\mathrm{D}(-\boldsymbol{k}),\\
&\check{C}_{+} \; \check{H}_\mathrm{D}(\boldsymbol{k}) \; \check{C}_{+}^{-1}
=\check{H}_\mathrm{D}(-\boldsymbol{k}),
\end{align}
with $\check{C}_+=\tau_1 {\cal K}$.
The chiral symmetry operator of $\check{H}(\boldsymbol{k})$ is then given by
\begin{align}
\check{S} = \check{T}_{+,1} \check{C}_+ = \left[ \begin{array}{cc}
0 & i \sigma_1 \\ - i \sigma_1 & 0 \\ \end{array}\right].
\end{align}

\begin{figure}[hhhh]
\begin{center}
\includegraphics[width=0.3\textwidth]{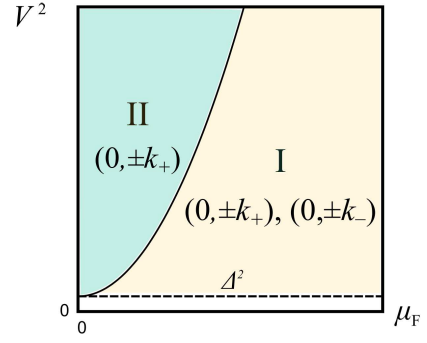}
\caption{(Color online)
Phase diagram of a Dresselhaus [110] superconductor.
The solid line represents $V^2=\mu_{\mathrm{F}}^2$.
The dashed line represents $V^2=\Delta_s^2$.
}
\label{fig:pd_d}
\end{center}
\end{figure}

The energy spectra of $\check{H}_\mathrm{D}(\boldsymbol{k})$ are calculated to be
\begin{align}
&E(\boldsymbol{k})=
\pm \sqrt{\varepsilon^2(\boldsymbol{k}) + \beta^2 k_x^2 + V^2 + \Delta_s^2 \pm 2 \eta(\boldsymbol{k})},\\
&\eta(\boldsymbol{k})=\sqrt{\varepsilon^2(\boldsymbol{k})\beta^2 k_x^2+V^2\left(\varepsilon^2(\boldsymbol{k})+\Delta_s^2 \right)},
\end{align}
where $V=\sqrt{V_1^2 + V_2^2}$ represents the amplitude of the Zeeman field.
A Dresselhaus[110] superconductor has two superconducting phases in terms of the number of point nodes 
on the Fermi surface: four point nodes in phase I and two point nodes in phase II.
The phase diagram is shown in Fig.~\ref{fig:pd_d}.
The phase I is characterized by the relation $\Delta_s^2<V^2<\mu_\mathrm{F}^2+\Delta_s^2$.
The nodal points are given by $(k_x,k_y)=(0,\pm k_+)$ and $(0,\pm k_-)$ with
\begin{align}
k_{\pm}=\frac{\sqrt{2m\left(\mu_\mathrm{F} \pm \sqrt{V^2-\Delta_s^2}\right)}}{\hbar}.
\end{align}
On the other hand, the phase II is characterized by $V^2 >\mu_\mathrm{F}^2+\Delta_s^2$.
The resulting nodal points are located at $(k_x,k_y)=(0,\pm k_+)$.

Now we focus on the flat-band ZESs appearing at the surface parallel to the $y$ direction.
The winding number in Eq.~(\ref{eq:wind}) can be further simplified to~\cite{sato_2}
\begin{align}
&w(k_y) = - \frac{1}{2} \sum_{k_x \, \text{at}\,  m_1(\boldsymbol{k}) =0}
\mathrm{sgn} [ \partial_{k_x} m_1(\boldsymbol{k}) ]
\mathrm{sgn} [ m_2 (\boldsymbol{k}) ], \label{eq:wind_dres}\\
&m_1(\boldsymbol{k})=\varepsilon^2(\boldsymbol{k}) - \beta^2 k_x^2 - V^2 + \Delta_s^2,\\
&m_2(\boldsymbol{k})=2 \beta k_x.
\end{align}
The summation is carried out for wave numbers in the $x$ direction $k_x$ that satisfy $m_1(\boldsymbol{k}) =0$ at a fixed $k_y$.
The results for the phase I are given by 
\begin{align}
w(k_y)=
\left\{ \begin{array}{cl} 
-1 & \text{for}\quad k_-<|k_y|< k_+ \\
0 & \text{otherwise},
\end{array}\right.
\label{eq:wind_d1}
\end{align}
and those for phase II are given by
\begin{align}
w(k_y)=
\left\{ \begin{array}{cl} 
-1 & \text{for}\quad 0 \leq |k_y|< k_+, \\
0 & \text{otherwise}.
\end{array}\right.
\label{eq:wind_d2}
\end{align}
The number of the zero-energy states at a dirty surface is evaluated by
the index ${\cal N}_\mathrm{ZES}$.
By substituting Eqs.~(\ref{eq:wind_d1}) and (\ref{eq:wind_d2}) into Eq.~(\ref{eq:n_zes}),
we obtain
\begin{align}
{\cal N}_\mathrm{ZES}=
\left\{ \begin{array}{cl} 
-\sum_{k_-<|k_y|< k_+} & \text{for phase I}, \\
- \sum_{0 \leq |k_y|< k_+} & \text{for phase II}.
\end{array}\right. \label{eq:nzes_d}
\end{align}
The index ${\cal N}_\mathrm{ZES}$ is nonzero in both phases.
Therefore,
$|{\cal N}_\mathrm{ZES}|$-fold degenerate ZESs are expected
at a dirty surface of the Dresselhaus [110] superconductor.

%------------------------------------------------------------------------
%------------------------------------------------------------------------

%%%%%%%%%%%%%%%%%%%%%%%%%%%%%%%%%%%%%%%%%%%%
\subsection{Helical $p$-wave superconductor with the in-plane magnetic field}
%%%%%%%%%%%%%%%%%%%%%%%%%%%%%%%%%%%%%%%%%%%%
The second example requires two-dimensional helical $p$-wave superconductivity. 
It is well known that a helical $p$-wave superconductor is fully gaped and hosts helical edge states 
at its surface reflecting a nonzero $\mathcal{Z}_2$ invariant. 
Here, we apply an in-plane magnetic field~\cite{sato_12, law}. 
The BdG Hamiltonian is described by
\begin{align}
&\check{H}_\mathrm{P}(\boldsymbol{k}) = \left[ \begin{array}{cc}
\hat{h}_\mathrm{P}(\boldsymbol{k}) &
\hat{\Delta}_\mathrm{P}(\boldsymbol{k}) \\
-\hat{\Delta}_\mathrm{P}^{\ast}(-\boldsymbol{k}) &
-\hat{h}_\mathrm{P}^{\ast} (-\boldsymbol{k}) \\
\end{array}\right], \label{eq:bdg_hp}\\
&\hat{h}_\mathrm{P}(\boldsymbol{k})=
\varepsilon(\boldsymbol{k}) \sigma_0 + \sum_{j=1,2} V_j \sigma_j \\
&\hat{\Delta}_\mathrm{P}(\boldsymbol{k}) =
i \frac{\Delta_p}{k_\mathrm{F}} \left[ k_x \hat{\sigma}_1 + k_y \hat{\sigma}_2 \right] \hat{\sigma}_2,
\end{align}
where $\Delta_p$ is the amplitude of the helical $p$-wave pair potential
and $k_\mathrm{F}=\sqrt{2m \mu_\mathrm{F}}/\hbar$ is the Fermi wave number.
The Zeeman potential breaks TRS and spin-rotation symmetry simultaneously.
Nevertheless, the BdG Hamiltonian is classified into class BDI~\cite{sato_2014}, where
\begin{align}
&\check{T}_{+,2} \; \check{H}_\mathrm{P}(\boldsymbol{k}) \; \check{T}_{+,2}^{-1}=
\check{H}_\mathrm{P}(-\boldsymbol{k}),\\
&\check{C}_{+} \; \check{H}_\mathrm{P}(\boldsymbol{k}) \; \check{C}_{+}^{-1}=
\check{H}_\mathrm{P}(-\boldsymbol{k}),
\end{align}
are satisfied.
The chiral symmetry operator is then given by
\begin{align}
\check{S} = \check{T}_{+,2}\check{C}_{+} = \left[ \begin{array}{cc}
0 & \sigma_1 \\ \sigma_1 & 0 \\ \end{array}\right].
\end{align}

\begin{figure}[hhhh]
\begin{center}
\includegraphics[width=0.3\textwidth]{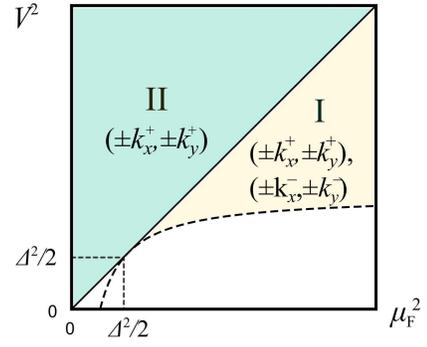}
\caption{(Color online)
Phase diagram of a helical $p$-wave superconductor under an in-plane Zeeman potential.
The solid line represents $V^2=\mu_{\mathrm{F}}^2$.
The dashed line represents $V^2=-(\Delta_p^4/4\mu_{\mathrm{F}}^2)+\Delta_p^2$.
}
\label{fig:pd_h}
\end{center}
\end{figure}

The energy eigenvalues of $\check{H}_\mathrm{P}(\boldsymbol{k})$ are calculated to be
\begin{align}
&E(\boldsymbol{k})=
\pm \sqrt{\varepsilon^2(\boldsymbol{k}) + V^2 + \Delta_p^2 k^2 \pm 2 \zeta(\boldsymbol{k})},\\
&\zeta(\boldsymbol{k})=\sqrt{\varepsilon^2(\boldsymbol{k})V^2 + \Delta_p^2 \left(V_1 k_x + V_2 k_y \right)^2 }.
\end{align}
A helical $p$-wave superconductor under an in-plane magnetic field has three superconducting phases.
The phase I appears when the parameters satisfy $-(\Delta_p^4/4\mu_{\mathrm{F}}^2)+\Delta_p^2
< V^2 < \mu_{\mathrm{F}}^2$ and $\mu_{\mathrm{F}}^2 > \Delta_p^2/2$.
The four nodal points 
on the Fermi surface are given by
$(k_x^+, k_y^+)$, $(-k_x^+, -k_y^+)$,$(k_x^-, k_y^-)$, and $(-k_x^-, -k_y^-)$ with
\begin{align}
&k_x^{\pm} = k_0^{\pm} \cos (\theta_V), \quad k_y^{\pm} = k_0^{\pm} \sin (\theta_V),\\
&k_0^{\pm} = \sqrt{k_\mathrm{F}^2 - 2 \kappa^2 \pm \sqrt{k_V^4 - 4 \kappa^2 \left( k_\mathrm{F}^2 - \kappa^2 \right)}},\\
&k_V=\frac{\sqrt{2mV}}{\hbar}, \quad \kappa = \frac{m \Delta_p}{\hbar^2 k_\mathrm{F}},
\quad \theta_V = \arctan \left( \frac{V_2}{V_1} \right).
\end{align}
In the phase II appearing at $V^2 > \mu_{\mathrm{F}}^2$, there are two nodal points at $(k_x^+, k_y^+)$ and $(-k_x^+,- k_y^+)$.
Finally, the superconducting states are topologically trivial in the lest of the parameter region.
The phase diagram is shown in Fig.~\ref{fig:pd_h}.

The winding number is calculated as
\begin{align}
&w(k_y) = - \frac{1}{2} \sum_{k_x\, \text{at}\, m^\prime_1(\boldsymbol{k}) =0}
\mathrm{sgn} [ \partial_{k_x} m^\prime_1(\boldsymbol{k}) ]
\mathrm{sgn} [ m^\prime_2 (\boldsymbol{k}) ],\\
&m^\prime_1(\boldsymbol{k})=\varepsilon^2(\boldsymbol{k}) - V^2 + \Delta_p^2 \boldsymbol{k}^2 ,\\
&m^\prime_2(\boldsymbol{k})=\Delta_p( V_1 k_y-V_2 k_x),
\end{align}
where the summation is carried out for $k_x$ satisfying $m_1(\boldsymbol{k}) =0$ at a fixed $k_y$.
The results are given by
\begin{align}
w(k_y)=
\left\{ \begin{array}{cl} 
s_V & \text{for}\quad |k_y^-| <|k_y|< |k_y^+|, \\
0 & \text{otherwise},
\end{array}\right. \label{eq:wind_h1}
\end{align}
for the phase I and 
\begin{align}
w(k_y)=
\left\{ \begin{array}{cl} 
s_V & \text{for}\quad |k_y|< |k_y^+|, \\
0 & \text{otherwise}.
\end{array}\right. \label{eq:wind_h2}
\end{align}
for the phase II with $s_V = \mathrm{sgn}[\sin \left(\theta_V\right)]$.
At $\theta_V=0$ or $\pi$, the winding number becomes zero for all $k_y$
because of $k_y^{\pm}=0$.
When $\theta_V$ is neither $0$ nor $\pi$, 
the winding number $w(k_y)$ can be either $+1$ or $-1$ depending on $s_V$.
By substituting Eqs.~(\ref{eq:wind_h1}) and (\ref{eq:wind_h2}) into Eq.~(\ref{eq:n_zes}), we obtain
\begin{align}
{\cal N}_\mathrm{ZES}=
\left\{ \begin{array}{cl} 
s_V \sum_{|k_y^-|<|k_y|< |k_y^+|} & \text{phase I}, \\
s_V \sum_{0 \leq |k_y|< |k_y^+|} & \text{phase II}.
\end{array}\right. \label{eq:nzes_hp}
\end{align}
The nonzero index ${\cal N}_\mathrm{ZES}$ in Eq.~(\ref{eq:nzes_hp})
suggest the existence of the stable $|{\cal N}_\mathrm{ZES}|$-fold degenerate ZESs
at a dirty surface of a helical $p$-wave superconductor under an in-plane magnetic field.

%------------------------------------------------------------------------
%------------------------------------------------------------------------

%%%%%%%%%%%%%%%%%%%%%%%%%%%%%%%%%%%%%%%%%%%%
\section{Conclusion}
\label{sec:4}
%%%%%%%%%%%%%%%%%%%%%%%%%%%%%%%%%%%%%%%%%%%%
We studied the symmetry property of a nodal superconductor
that hosts robust flat-band zero-energy states (ZESs) at its dirty surface.
A nodal superconductor is topologically characterized by 
the winding number defined in a one-dimensional partial Brillouin zone. 
On the basis of the bulk boundary correspondence,
we show the existence of flat-band ZESs at the clean surface of a nodal superconductor
belonging to any of the symmetry classes BDI, CII, DIII, or CI. 
In the presence of potential disorder,  
we find that surface flat-band ZESs are robust only when the nodal superconductor belongs to either class BDI or class CII.
In addition, we investigated two realistic examples of single-band nodal superconductors that belong to
class BDI: a Dresselhaus [110] superconductor and a helical $p$-wave superconductor under a magnetic field.
We found that flat-band ZESs are stable at a dirty surface in both cases.
Therefore, such superconductors are promising candidates for observing the anomalous proximity effect, 
which is drastic phenomena caused by flat-band ZESs.

\begin{acknowledgments}
This work was supported by ``Topological Materials Science'' (No. JP15H05852)
and KAKENHI (Nos. JP26287069 and JP15H03525)
from the Ministry of Education, Culture, Sports, Science and Technology (MEXT)
of Japan and by the Ministry of Education and Science of
the Russian Federation (Grant No. 14Y.26.31.0007).
 SI is supported in part by a Grant-in-Aid for JSPS
Fellows (Grant No. JP16J00956) provided by the Japan Society for the Promotion of Science (JSPS).
SK is supported by the Grant-in-Aid for Scientific Research B (Grant No. JP17H02922),
the Grant-in-Aid for Research Activity Start-up (Grant No. JP16H06861),
and the Building of Consortia for the Development of Human Resources in Science and Technology.
\end{acknowledgments}

\appendix
%%%%%%%%%%%%%%%%%%%%%%%%%%%%%%%%%%%%%%%%%%%%
\section{Commutation relations of symmetry operators}
\label{sec:com}
%%%%%%%%%%%%%%%%%%%%%%%%%%%%%%%%%%%%%%%%%%%%

We summarize the commutation relation among the symmetry operators.
The Hamiltonian under consideration preserves
time-reversal symmetry (TRS) and particle-hole symmetry (PHS) as
\begin{align}
&{\cal T} \,{\cal H} (\boldsymbol{k})\, {\cal T}^{-1} = {\cal H} (-\boldsymbol{k}),
\quad {\cal T}^2 = \eta_{\cal T},
\quad \eta_{\cal T} = \pm 1,\\
&{\cal C} \, {\cal H} (\boldsymbol{k})\,  {\cal C}^{-1} = - {\cal H} (-\boldsymbol{k}),
\quad {\cal C}^2 = \eta_{\cal C},
\quad \eta_{\cal C} = \pm 1,
\end{align}
where ${\cal T}$ and ${\cal C}$ are anti-unitary operator.
By combining TRS and PHS, the Hamiltonian also preserves chiral symmetry (CS) as
\begin{align}
{\cal S}\, {\cal H} (\boldsymbol{k})\, {\cal S}^{-1} = - {\cal H} (\boldsymbol{k}),
\quad
{\cal S} = e^{i \alpha_0}\, {\cal T}\, {\cal C},
\quad
{\cal S}^2=+1
\label{eq:cs2}
\end{align}
where $\alpha_0$ is determined so that ${\cal S}^2=+1$ is satisfied.
Since ${\cal T}\, {\cal C}\, {\cal C}\, {\cal T} = \eta_{\cal T}\, \eta_{\cal C}$,
we immediately find ${\cal T}\, {\cal C} = \eta_{\cal T}\, \eta_{\cal C}\, {\cal T} ^{-1}{\cal C}^{-1}$.
This leads the relation
\begin{align}
({\cal T}{\cal C})^2 = \eta_{\cal T}\, \eta_{\cal C}\, {\cal T} \,{\cal C}\, {\cal T}^{-1}{\cal C}^{-1}.
\label{eq:tc1}
\end{align}
From Eq.~(\ref{eq:cs2}), we also obtain
\begin{align}
{\cal S}^2 = e^{2 i \alpha_0} ({\cal T}{\cal C})^2 = + 1
\label{eq:tc2}
\end{align}
From Eqs.~(\ref{eq:tc1}) and (\ref{eq:tc2}), we find the relation
$\eta_{\cal T}\, \eta_{\cal C}\, {\cal T}\, {\cal C} {\cal T}^{-1}{\cal C}^{-1} = e^{- 2 i \alpha_0}$,
which can be deformed as
\begin{align}
{\cal C}{\cal T}   = e^{2 i \alpha_0}\, \eta_{\cal T}\, \eta_{\cal C}\, {\cal T}\, {\cal C}.
\label{eq:tc3}
\end{align}
By using Eq.~(\ref{eq:tc3}), we obtain
\begin{align}
{\cal S}{\cal T} &= e^{i \alpha_0} {\cal T}{\cal C}{\cal T}
= {\cal T} e^{- i \alpha_0} \left( e^{2 i \alpha_0}\, \eta_{\cal T}\, \eta_{\cal C}\, {\cal T}\, {\cal C} \right) \nonumber\\
&= \eta_{\cal T}\, \eta_{\cal C}\, {\cal T}\,{\cal S}, \\
{\cal S}{\cal C} &= e^{i \alpha_0} {\cal T}{\cal C}{\cal C}
= e^{i \alpha_0} \left( e^{- 2 i \alpha_0}\, \eta_{\cal T}\, \eta_{\cal C}\, {\cal C}\, {\cal T} \right){\cal C} \nonumber\\
&= \eta_{\cal T}\,  \eta_{\cal C}\,  {\cal C}\, {\cal S}.
\end{align}
As a consequence, we find the commutation relation
\begin{align}
[{\cal S}, {\cal T}] = [{\cal S}, {\cal C}] = 0,
\end{align}
for $\eta_{\cal T}\, \eta_{\cal C} =+1$, and
\begin{align}
\left\{{\cal S}, {\cal T}\right\} = \left\{{\cal S}, {\cal T}\right\} = 0,
\end{align}
for $\eta_{\cal T}\, \eta_{\cal C} =-1$.

%%%%%%%%%%%%%%%%%%%%%%%%%%%%%%%%%%%%%%%%%%%%
\section{Anti-unitary operator $\boldsymbol{\check{T}_+}$}
\label{sec:aut}
%%%%%%%%%%%%%%%%%%%%%%%%%%%%%%%%%%%%%%%%%%%%
We explain the expression of the anti-unitary operator $\check{T}_+$ which satisfies $\check{T}_+^2=+1$.
By combining $\check{T}_-$ and an unitary operator $\check{R}$, it is possible to define $\check{T}_+$ as
\begin{align}
&\check{T}_+ = \check{R} \; \check{T}_-
= \left[ \begin{array}{cc}
\hat{r} \; \hat{T}_-  & 0 \\ 0 & \hat{r}^{\ast} \hat{T}_- \\ \end{array}\right], \\
&\check{R}= \left[ \begin{array}{cc}
\hat{r}  & 0 \\ 0 &  \hat{r}^{\ast}  \\ \end{array}\right], \quad
\check{T}_-= \left[ \begin{array}{cc}
\hat{T}_-  & 0 \\ 0 &  \hat{T}_-  \\ \end{array}\right] , \quad
\hat{T}_- = i \sigma_2 {\cal K}
\end{align}
where $\hat{r}$ is a $2 \times 2$ unitary operator and $\check{T}_-^2 = -1$.
The unitary operator $\hat{r}$ must satisfies
\begin{align}
\left( \hat{r} \hat{T}_- \right)^2 = +1,
\label{eq:rtrt}
\end{align}
so that the relation $\check{T}_+^2=+1$ holds.

A general expression of a $2 \times 2$ unitary operator is given by
\begin{align}
\hat{r} &=  e^{i \gamma/2}\hat{r}_0,\label{eq:abr}\\
\hat{r}_0 &=
\mathrm{exp} \left[- i \frac{\phi}{2} \boldsymbol{n} \cdot \boldsymbol{\sigma} \right]
\nonumber\\
&=\left[
\cos \left( \frac{\phi}{2} \right) \sigma_0
- i \sin \left( \frac{\phi}{2} \right) \boldsymbol{n} \cdot \boldsymbol{\sigma} \right],
\end{align}
where $\gamma$ and $\phi$ are arbitrary real numbers
and $\boldsymbol{n}$ is a unit vector in an arbitrary direction.
By substituting Eq.~(\ref{eq:abr}) into Eq.~(\ref{eq:rtrt}), we obtain
\begin{align}
\left( \hat{r} \; \hat{T}_- \right)^2
&= \hat{r} \; ( i \sigma_2) \; \hat{r}^{\ast} \; ( i \sigma_2) \nonumber\\
&= - \hat{r}_0 \; \sigma_2 \; \hat{r}_0^{\ast} \; \sigma_2 \nonumber\\
&= - \hat{r}_0^2 \nonumber\\
&= - \left[  \cos^2\left( \frac{\phi}{2} \right)
- i \sin (\phi) \boldsymbol{n} \cdot \boldsymbol{\sigma}
- \sin^2\left( \frac{\phi}{2} \right) \right] \nonumber\\
& = +1. \label{eq:rtrt2}
\end{align}
The equation (\ref{eq:rtrt2}) is satisfied only when $\phi = \pm \pi$.
Therefore, the unitary operator $\hat{r}$ is restricted to
\begin{align}
\hat{r} = - i \, e^{i \gamma/2} \boldsymbol{n} \cdot \boldsymbol{\sigma},
\end{align}
to satisfy $\check{T}_+^2=+1$.

%%%%%%%%%%%%%%%%%%%%%%%%%%%%%%%%%%%%%%%%%%%%
\section{Unitary transformation}
\label{sec:uni}
%%%%%%%%%%%%%%%%%%%%%%%%%%%%%%%%%%%%%%%%%%%%
We consider the BdG Hamiltonian preserving time-reversal symmetry (TRS) as
\begin{align}
&\check{T}_+ \; \check{H}(\boldsymbol{k}) \; \check{T}_+^{-1}=\check{H}(-\boldsymbol{k}), \label{apc:trsp}\\
&\check{T}_+ = \left[ \begin{array}{cc}
\hat{r}( i \sigma_2 ) & 0 \\ 0 & \hat{r}^{\ast} ( i \sigma_2 ) \\ \end{array}\right] {\cal K},
\end{align}
where the $2 \times 2$ unitary operator $\hat{r}$ is given by
\begin{align}
&\hat{r} = - i e^{i \gamma/2} \boldsymbol{n} \cdot \boldsymbol{\sigma}, \\
&\boldsymbol{n} = ( \cos \varphi \sin \theta, \sin \varphi  \sin \theta, \cos \theta ).
\end{align}
When we apply a unitary transformation as
\begin{align}
&\check{H}_z(\boldsymbol{k}) =
\check{U} \; \check{H}(\boldsymbol{k}) \; \check{U}^{\dagger}, \quad
\check{T}_z(\boldsymbol{k}) =
\check{U} \; \check{T}_+ \; \check{U}^{\dagger},
\end{align}
with
\begin{align}
&\check{U} = \left[ \begin{array}{cc}
\hat{u}  & 0 \\ 0 &  \hat{u}^{\ast}  \\ \end{array}\right] , \\
&\hat{u}= \left[ \begin{array}{cc}
e^{i \varphi/2} \cos \left(\frac{\theta}{2} \right)  & e^{- i \varphi/2} \sin \left(\frac{\theta}{2} \right) \\
- e^{i \varphi/2} \sin \left(\frac{\theta}{2} \right) &  e^{- i \varphi/2} \cos \left(\frac{\theta}{2} \right)  \\ \end{array}\right],
\end{align}
TRS of $\check{H}_z(\boldsymbol{k})$ is represented by
\begin{align}
&\check{T}_z \; \check{H}_z(\boldsymbol{k}) \; \check{T}_z^{-1} =\check{H}_z(-\boldsymbol{k}),
\label{eq:t+_app}\\
&\check{T}_z =  \check{R}_z \; \check{T}_-
= \left[ \begin{array}{cc}
-i e^{i \gamma/2} \sigma_1  & 0 \\ 0 & i e^{-i \gamma/2} \sigma_1  \\ \end{array}\right] {\cal K},\\
&\check{R}_z= \left[ \begin{array}{cc}
\hat{r}_z  & 0 \\ 0 &  \hat{r}_z^{\ast}  \\ \end{array}\right], \quad
\hat{r}_z= - i e^{i \gamma/2} \sigma_3.
\end{align}
The results suggest that a BdG Hamiltonian preserving TRS in Eq.~(\ref{apc:trsp})
is always unitary equivalent to another BdG Hamiltonian preserving TRS in Eq.~(\ref{eq:t+_app}).


\begin{thebibliography}{32}
\bibitem{kane_1} M. Z. Hasan and C. L. Kane, Rev. Mod. Phys. \textbf{82}, 3045 (2010).
\bibitem{shou}  X~-L.~Qi  and  S.~C~Zhang, Rev. Mod. Phys. \textbf{83}, 1058 (2011).
\bibitem{sato_r} M. Sato and Y. Ando, Rep. Prog. Phys. \textbf{80}, 076501 (2017).
\bibitem{schnyder_0} A.~P.~Schnyder, S.~Ryu, A. Furusaki, A. W. W. Ludwig, Phys. Rev. B \textbf{78}, 195125 (2008).
\bibitem{yt95} Y.~Tanaka and S.~Kashiwaya, Phys. Rev. Lett. \textbf{74}, 3451 (1995). 
\bibitem{ya04} Y.~Asano, Y.~Tanaka, and S.~Kashiwaya, Phys. Rev. B \textbf{69}, 134501 (2004). 
\bibitem{sato_2} M.~Sato, Y.~Tanaka, K.~Yada, and T.~Yokoyama, Phys. Rev. B \textbf{83}, 224511 (2011).
\bibitem{schnyder_5}A.~P.~Schnyder and P.~M.~R.~Brydon, J. Phys.: Condens. Matter \textbf{27}, 243201 (2015).
\bibitem{buchholts} L.~J.~Buchholtz and G.~Zwicknagl, Phys. Rev. B \textbf{23},~5788 ~(1981). 
\bibitem{hu} C.~R.~Hu,  Phys. Rev. Lett. \textbf{72}, 1526 (1994).
\bibitem{yt10} Y. Tanaka, Y.~Mizuno, T.~Yokoyama, K.~Yada, and M.~Sato, Phys. Rev. Lett. \textbf{105}, 097002 (2010).
\bibitem{yada} K.~Yada, M.~Sato, Y.~Tanaka, and T.~Yokoyama, Phys. Rev. B \textbf{83}, 064505 (2011).
\bibitem{schnyder_1} P.~M.~R.~Brydon, A.~P.~Schnyder,  and  C.~Timm, Phys. Rev. B \textbf{84}, 020501(R) (2011).
\bibitem{schnyder_2} A.~P.~Schnyder and  S.~Ryu, Phys. Rev. B \textbf{84}, 060504(R) (2011).
\bibitem{schnyder_3} A.~P.~Schnyder, P.~M.~R.~Brydon and  C.~Timm, Phys. Rev. B \textbf{85}, 024522 (2012).
\bibitem{alicea} J.~Alicea, Phys. Rev. B {\bf 81}, 125318 (2010).
\bibitem{you} J.~You, C.~H.~Oh, and V.~Vedral, Phys. Rev. B {\bf 87}, 054501 (2013).
\bibitem{si15} S.~Ikegaya, Y.~Asano, and Y.~Tanaka, Phys. Rev. B \textbf{91}, 174511 (2015).
\bibitem{sato_12} T. Mizushima, M. Sato and K. Machida, Phys. Rev. Lett. {\bf 109}, 165301(2012).
\bibitem{law} C.~L. M. Wong, J.~Liu, K.~T.~Law, P.~A.~Lee, Phys. Rev. B {\bf 88}, 060504(R) (2013).
\bibitem{deng} S.~Deng, G.~ Ortiz,  A.~Poudel, and L.~Viola, Phys. Rev. B \textbf{89}, 140507(R) (2014).
\bibitem{franz} A.~Chen and M.~Franz, Phys. Rev. B \textbf{93}, 201105(R) (2016).
\bibitem{kobayashi_2014} S. Kobayashi, K. Shiozaki, Y. Tanaka, and M. Sato, Phys. Rev. B \textbf{90}, 024516 (2014).
\bibitem{kobayashi_2015} S. Kobayashi, Y. Tanaka, and M. Sato, Phys. Rev. B \textbf{92}, 214514 (2015).
\bibitem{yt04} Y.~Tanaka and S.~Kashiwaya, Phys.~Rev.~B \textbf{70}, 012507 (2004).
\bibitem{ya07} Y.~Asano, Y.~Tanaka, A.~A.~Golubov, and S. Kashiwaya, Phys.~Rev.~Lett. \textbf{99}, 067005 (2007).
\bibitem{si16} S. Ikegaya, S.-I. Suzuki,  Y. Tanaka, and Y. Asano, Phys.\ Rev.~B \textbf{94}, 054512 (2016).
\bibitem{yt96-1} Y. Tanaka and S. Kashiwaya, Phys. Rev. B \textbf{53}, 11957(R) (1996).
\bibitem{barash} Y. S. Barash, H. Burkhardt and D. Rainer, Phys. Rev. Lett. \textbf{77}, 4070 (1996).
\bibitem{kwon} H. -J. Kwon, K. Sengupta, and V. M. Yakovenko,  Eur.\ Phys.\ J.~B \textbf{37}, 349$-$361(2004).
\bibitem{ya06} Y.\ Asano, Y.\ Tanaka, and S.\ Kashiwaya, Phys.\ Rev.\ Lett.\ \textbf{96}, 097007 (2006).
\bibitem{si16_2}S. Ikegaya and Y. Asano, J. Phys.: Condens. Matter \textbf{28}, 375702 (2016).
\bibitem{higashitani}  S.\ Higashitani, J. Phys. Soc. Jpn., \textbf{66}, 2556 (1997).
\bibitem{walter} 
H.~Walter, W.~Prusseit, R.~Semerad, H.~Kinder, W.~Assmann, H.~Huber,
H.~Burkhardt, D.~Rainer, and J.~A.~Sauls,  Phys. Rev. Lett. \textbf{80}, 3598 (1998). 
\bibitem{vorontsov} A.\ B.\ Vorontsov, Phys. Rev. Lett. \textbf{102}, 177001 (2009).
\bibitem{suzuki1} S.-I.~Suzuki and Y.~Asano, Phys. Rev. B \textbf{89}, 184508 (2014).
\bibitem{suzuki2} S.-I.~Suzuki and Y.~Asano, Phys. Rev. B \textbf{91}, 214510 (2015). 
\bibitem{fogelstrom} M.\ Hakansson, T.\ Lofwander and M.\ Fogelstrom, Nat. Phys. \textbf{11}, 755 (2015).
\bibitem{taddei_15} S. Valentini, R. Fazio, V. Giovannetti, and F. Taddei, Phys. Rev. B \textbf{91}, 045430 (2015)
\bibitem{ya01} Y.~Asano, Phys. Rev. B \textbf{63}, 052512 (2001).
\bibitem{yt03} Y.~Tanaka, Yu.~V. Nazarov, and S.~Kashiwaya, Phys. Rev. Lett. \textbf{90}, 167003 (2003). 
\bibitem{schnyder_6}R. Queiroz an A.~P.~Schnyder, Phys. Rev. B \textbf{89}, 054501 (2014).
\bibitem{si17} S. Ikegaya and Y. Asano, Phys.\ Rev.~B \textbf{95}, 214503 (2017).
\bibitem{yamakage} Y.~Xiong, A.~Yamakage, S.~Kobayashi, M.~Sato, and Y.~Tanaka, Crystals \textbf{7}, 58 (2017).
\bibitem{sato_2014} K. Shiozaki and M. Sato, Phys.\ Rev.~B \textbf{90}, 165114 (2014).
\end{thebibliography}
\end{document}